\documentclass[10pt, conference]{IEEEtran}

\usepackage[utf8]{inputenc}
\usepackage[T1]{fontenc}
\usepackage{amsmath,amssymb,amsthm}
\usepackage{graphicx}
\usepackage{hyperref}
\usepackage{cite}
\usepackage{geometry}
\usepackage{braket}
\usepackage{booktabs}
\usepackage{algorithm}
\usepackage{algpseudocode}

\usepackage{tikz}
\usetikzlibrary{arrows.meta,calc,fit,backgrounds,shapes.geometric}
 
\definecolor{leafc}{RGB}{225,245,238}
\definecolor{prec}{RGB}{238,237,254}
\definecolor{quec}{RGB}{241,239,232}
\definecolor{tdc}{RGB}{253,237,232}
 
\tikzset{
  q/.style={draw, rounded corners=2pt, fill=quec, align=center,
            text width=5.6cm, minimum height=1.1cm, font=\footnotesize},
  leaf/.style={draw, rounded corners=2pt, fill=leafc, align=center,
               text width=5.6cm, minimum height=1.1cm, font=\footnotesize},
  pre/.style={draw, rounded corners=2pt, fill=prec, align=center,
              text width=5.6cm, minimum height=1.1cm, font=\footnotesize},
  decision/.style={draw, diamond, aspect=1.6, fill=tdc, align=center,
                   text width=2.3cm, inner sep=1pt, font=\scriptsize},
  tdleaf/.style={draw, rounded corners=2pt, fill=leafc, align=center,
                 text width=4.4cm, minimum height=1.1cm, font=\footnotesize},
  flow/.style={-{Stealth[length=5pt]}, thin},
  lab/.style={font=\scriptsize, inner sep=2pt}
}

\newtheorem{theorem}{Theorem}[section]

\newtheorem{corollary}{Corollary}[section]

\newcommand{\norm}[1]{\left\|#1\right\|}

\title{Product Formulas for Hamiltonian Simulation: A Survey of Optimization Techniques}
\author{
    \IEEEauthorblockN{Anna Matsekh, John Golden, Reuben Tate, Stephan Eidenbenz}
    \IEEEauthorblockA{
        \textit{Los Alamos National Laboratory}\\
        Email: \{matsekh, golden, rtate, eidenben\}@lanl.gov
    }
}

\begin{document}

\maketitle

\begin{abstract}
Product formula methods for Hamiltonian simulation, or Trotterization, have long become a standard approximation approach in quantum computing. Based on the fundamental Lie-Trotter-Suzuki formulas, Trotterization remains an active area of quantum algorithm research, even in the presence of highly competitive alternatives, such as the quantum singular value transformation, qubitization, and truncated Taylor series based methods. Trotterization often depends on the structure of the Hamiltonian operator, and rigorous worst-case error bounds can be overly pessimistic. However,  with relatively well understood algebraic properties and minimal ancilla qubit overheads, product formulas are widely used in Hamiltonian algorithm design and research targeting near term quantum architectures. The goal of this paper is to survey a variety of distinct Trotterization methods in the context of the implementation optimality. We begin this paper with a historical note, followed by a brief summary of the most significant Trotter error analysis results, including commutator-based error bounds, lower bounds, as well as the error bounds in application specific instances. We next survey various methods for analyzing both application-specific and numerical, application agnostic Trotterization methods, including term ordering, randomized and partially randomized formulas, multi-product formulas, adaptive Trotterization, and methods that exploit multiple energy scales.  
\end{abstract}


\section{Introduction}

Product formula methods, or Trotterization, use numerical approximations to solve a hard to compute problem, that involves an exponential operator, as an easier to compute, rapidly convergent, product of exponential terms. The original Lie-Trotter product formula is defined for two arbitrary bounded non-commuting operators $A$ and $B$, and a parameter $x$ as follows~\cite{Hatano2005}
\begin{equation}
e^{x (A+B)} = e^{x A} e^{x B} + O(x^2), 
    \label{eq:lie-trotter}
\end{equation}
or, equivalently, as
\begin{equation}
e^{x A} e^{x B} = e^{x (A + B) + O(x^2)}. 
    \label{eq:lie-trotter-1}
\end{equation}
The Lie-Trotter formula~\eqref{eq:lie-trotter} is the first order accurate in $x$, and forms a basis of numerous rapidly convergent higher order modern product formulas, often used in computational physics, chemistry, engineering, and, most recently, in quantum computing.  Product formulas are based on the fundamental 1888 results of Lie~\cite{Lie-1888}, Trotter's 1958-1959 research~\cite{Trotter-58, Trotter-1959}, and the works of Suzuki~\cite{suzuki-75, suzuku-85, suzuki-92, Hatano2005}, spanning decades between the 1970-s and the 2000-s. Trotterization has become a standard numerical technique in quantum computing used to implement unitary time evolution 
\begin{equation}
U(\tau)=  e^{-i H \tau}
\end{equation} 
under a Hamiltonian operator $H$. Trotterization is a natural approximation approach for many physical systems,
such as the Hubbard and Ising models, which are represented as sums over many local interactions~\cite{Nielsen-Chuang}
\begin{equation}
H = \sum_{k=1}^{L} H_k. 
\end{equation}
Many significant contributors to the modern Trotterization theory have been made by Suzuki~\cite{suzuki-75, suzuku-85, suzuki-92,Hatano2005}. Suzuki's first order accurate Generalized Trotter formula~\cite{suzuki-75}
\begin{equation}
S_1(\tau) = \prod_{k=1}^L e^{-iH_k \tau} = e^{-iH_L\tau} \cdots e^{-iH_2\tau} e^{-iH_1\tau}
\label{gen-trotter}
\end{equation}
is defined for an arbitrary number of bounded operators $H_k$. Suzuki derived the Generalized Trotter formula~\eqref{gen-trotter} by proposing a rapidly convergent approximation of the exponential function, that he extended  to define an approximation of the exponential operator~\eqref{gen-trotter}. Represented by a $2^n \times 2^n$ Hermitian matrix in Hilbert space, a Hamiltonian operator acts on n qubits on a quantum devise, and while the unitary evolution operator $U(\tau)$ is generally difficult to compute, each term $e^{-i H_k \tau}$ acts on a much smaller subsystem, and can be implemented as a quantum circuit~\cite{Nielsen-Chuang}. 

The theory of operator splitting and the use of fractional steps is deeply rooted in the approximation methods of computational physics.  Indeed,  Wei, Guenter, and Spiteri~\cite{wei-guenter-25},~\cite{Spiteri-Wei-23}, note that in the computational physics community the Lie-Trotter formula is also known as Godunov's splitting (the first order), based on S.K.\,Godunov's 1958-1959 difference scheme for the solution of the equations of hydrodynamics~\cite{godunov-59}. The second order accurate Strang product formula is based on the 1968 work of Strang~\cite{strang-68}, who proposed the second order operator splitting
\begin{equation}
S_2(\tau) = \prod_{j=1}^m e^{-iH_j \tau/2} \prod_{j=m}^1 e^{-iH_j \tau/2}.
\end{equation}
Wei, Guenter, and Spiteri~\cite{wei-guenter-25},~\cite{Spiteri-Wei-23}, also note that Strang formula is often referred to as the second order accurate Strang–Marchuk splitting method. In turn, Strang–Marchuk formula is sometimes called Yanenko-Marchuk method of fractional steps, referring to the 1960s and 1970s works of N.N.\,Yanenko and G.I.\,Marchuk on a splitting method  for integral operators arising in computational physics~\cite{yanenko-91, yanenko-2012}. Furthermore, in a 1966 paper on the methods of splitting (fractional steps), Yanenko and Marchuk~\cite{yanenko-marchuk-66, yanenko-91} note that it was the 1955 work of Peaceman and Rachford~\cite{Peaceman-Rachford-55} on the Alternating-Direction Implicit Method that first used fractional steps and non-homogeneous approximations, paving the road for many important discoveries in modern approximation theory. 

The error of the $2 p$-th order Trotter-Suzuki approximation scales as  $O(\tau^{2 p + 1}/r^{2 p})$, where $r$ is the number of Trotter steps. While increasing the number of Trotter time steps improves algebraic accuracy of the approximation, it leads to increased quantum circuit complexity. As a consequence, the goal of attaining an optimal balance between the Trotterization error and optimal quantum circuit design has motivated a vast body of modern Trotterization research, focused on optimizing product formulas for application specific scenarios, striving to design optimal analytical representations, and numerical approximation schemes. 
Active Trotterization research is focused on such problems as application of higher order product formulas with further improved algebraic accuracy with trade-offs in the increased sampling and hardware dependent errors. Research is being conducted to identify applications where lower order approximations might be sufficient to achieve desired accuracy. Various novel numerical schemes are being developed, such as the randomized, partially randomized, adaptive, and multi-product formulas.  

The rest of the paper surveys these Trotterization algorithms and techniques, beginning with the section on error-bound analysis of the higher-order product formulas and lower error bounds, graph-based analysis of the commutator structure, analysis of application-specific quantum localization and error suppression techniques, as well as the error analysis in low-energy subspace simulations. The third section examines general numerical optimization techniques: graph-based term ordering, randomized and partially randomized formulas, adaptive Trotterization and multi-product formulas, as well as the state-of-the-art product formulas for Hamiltonians with multiple energy scales.

\section{Trotter Error Analysis}

\subsection{Higher-Order Product Formulas}

Relying on the general theory of fractal path integrals, Suzuki~\cite{Suzuki1991} introduced the following recursive definition of higher order product formulas, aiming to improve accuracy of the numerical approximation accordingly. Suzuki's~\cite{Suzuki1991} first- and second- order symmetrized product formulas have the following form:
\begin{equation}
S(x) = \prod_{k=1}^{q} e^{\frac{x}{2} A_k}\, e^{x A^{q+1}}\, \prod_{k=q}^{1} e^{\frac{x}{2} A_k}
\end{equation}
and, in general, 
\begin{equation}
\begin{split}
S_{2m}(x) &= S_{2m-1}(x) =  \\
&= (S_{2m-3}(p_m x))^2 \, S_{2m-3}((1 - 4 p_m) x) \\
&\times (S_{2m-3}(p_m x))^2,
\end{split}
\end{equation}
where parameter $p_m$ is the real solution of the equation
\begin{equation}
    4 p_m^{2 m -1} + (1 - 4 p_m)^{2 m -1} = 0,
\end{equation}
or, $p_m = (4 - 4^\frac{1}{(2m -1)})^{-1}$.  

Higher order product formulas improve numerical approximation error on near term quantum systems at the expense of producing complex quantum circuits that suffer from increased statistical sampling errors and hardware error accumulation. The problem of finding an optimal balance between the approximation accuracy and hardware error accumulation motivates the vast body of moderm Trotterization algorithm reseacrh. For instance, Ostmeyer~\cite{Ostmeyer_2023} provides a comprehensive analysis of recent optimized decomposition schemes both with real and complex coefficients, suggesting theoretically efficient unitary and non-unitary fourth-order decompositions, with computational schemes up to the eighth order, demonstrating that for many applications carefully chosen coefficients can reduce gate counts by two-to-five fold factors, compared to naive computational schemes.

\subsection{Commutator Structure}

The non-commutativity of Hamiltonian terms is the primary error source in product formulas. Using Baker-Campbell-Hausdorff (BCH) formula
\begin{equation}
e^A\, e^B = e^{A+B+\frac{1}{2}[A,B]+\frac{1}{12}[A,[A,B]]-\frac{1}{12}[B,[A,B]]+\cdots}
\end{equation}
Trotter error can be systematically expressed in terms of nested commutators~\cite{Hall-LieAlgebras}. Childs et al.~\cite{PhysRevX.11.011020} have developed a comprehensive theory of Trotter error based on truncating Baker-Campbell-Hausdorﬀ expansion, exploiting commutativity of operator summands and deriving  tighter error bounds for real and imaginary time evolution for general, non-local, Hamiltonian systems.  Childs et al.~\cite{PhysRevX.11.011020} were able to  reproduce known tight bounds for the first and the second order product formulas, and to demonstrate that developed analysis can accurately characterize Trotter
error in terms of asymptotic scaling and a constant prefactor. Childs et al. represent Trotter error in terms of nested commutators of the operator summands. Specifically, they provide upper bounds on the additive and multiplicative errors of the pth-order product formulas as follows: 
\begin{theorem}[Trotter error with commutator scaling]
\label{thm:11}
Let $H = \sum_{\gamma=1}^{\Gamma} H_\gamma$ be an operator consisting of $\Gamma$
summands and $t \ge 0$. Let
$\mathcal{S}(t) = \prod_{\upsilon=1}^{\Upsilon}\prod_{\gamma=1}^{\Gamma}
e^{t\,a_{(\upsilon,\gamma)}H_{\pi_\upsilon(\gamma)}}$
be a $p$th-order product formula. Define
\[
\widetilde{\alpha}_{\mathrm{comm}} \;=\;
\sum_{\gamma_1,\gamma_2,\dots,\gamma_{p+1}=1}^{\Gamma}
\bigl\| [H_{\gamma_{p+1}}, \cdots [H_{\gamma_2}, H_{\gamma_1}]] \bigr\|.
\]
Then, additive Trotter error and multiplicative Trotter error, defined
respectively by $\mathcal{S}(t) = e^{tH} + \mathcal{A}(t)$ and
$\mathcal{S}(t) = e^{tH}(I + \mathcal{M}(t))$, can be asymptotically bounded as
\begin{equation}
\begin{split}
\|\mathcal{A}(t)\|, \|\mathcal{M}(t)\|= O\!\left( \widetilde{\alpha}_{\mathrm{comm}}\, t^{p+1}
e^{\,4t\Upsilon \sum_{\gamma=1}^{\Gamma}\|H_\gamma\|} \right).
\end{split}
\end{equation}
Furthermore, if the $H_\gamma$ are anti-Hermitian, corresponding to physical Hamiltonians, 
\begin{equation}
\begin{split}
\|\mathcal{A}(t)\|, \|\mathcal{M}(t)\|  = O\!\left( \widetilde{\alpha}_{\mathrm{comm}}\, t^{p+1} \right).
\end{split}
\end{equation}
\end{theorem}

\begin{corollary}[Trotter number with commutator scaling]
\label{cor:12}
Let $H = \sum_{\gamma=1}^{\Gamma} H_\gamma$ be an operator consisting of $\Gamma$
summands with $H_\gamma$ anti-Hermitian and $t \ge 0$. Let
$\mathcal{S}(t) = \prod_{\upsilon=1}^{\Upsilon}\prod_{\gamma=1}^{\Gamma}
e^{t\,a_{(\upsilon,\gamma)}H_{\pi_\upsilon(\gamma)}}$
be a $p$th-order product formula. Define
\[
\widetilde{\alpha}_{\mathrm{comm}} \;=\;
\sum_{\gamma_1,\gamma_2,\dots,\gamma_{p+1}=1}^{\Gamma}
\bigl\| [H_{\gamma_{p+1}}, \cdots [H_{\gamma_2}, H_{\gamma_1}]] \bigr\|.
\]
Then, $\| \mathcal{S}^r(t/r) - e^{tH} \| = O(\varepsilon)$, provided that
\begin{equation}
r = O\!\left( \frac{ \widetilde{\alpha}_{\mathrm{comm}}^{1/p}\, t^{1+1/p} }
{ \varepsilon^{1/p} } \right).
\end{equation}
\end{corollary}

\subsection{Lower Bounds on Trotter Error}
Despite the tighter error bounds of Childs et al.~\cite{PhysRevX.11.011020}, upper error bounds often continue to overestimate actual simulation errors in many practical instances, e.g.~\cite{sahinoglu2021}. Hahn et al.~\cite{Hahn2025} provide explicit lower bounds on the Trotter error with the goal of finding optimistic error bounds that would help to  minimize quantum algorithm resource requirements. One of the two main results presented by 
Hahn et al.~\cite{Hahn2025} considers two normalized eigenstates $\varphi$ and $\psi$ of the Hamiltonian $A+B$, i.e., $(A+B)\varphi = h\varphi$ and $(A+B)\psi = \kappa\psi$, with $h \neq \kappa$, and spectral gap $\lambda \equiv |h-\kappa|$. Hahn et al.~\cite{Hahn2025} show that for all $t \geq 0$,  state-dependent Trotter error $\xi_n(t;\varphi)$ can be bounded from
below by
\begin{equation}
\xi_n(t;\varphi) \geq \max\{0, \sup_{\psi} \alpha_n(t;\varphi,\psi)\},
\label{eq:5}
\end{equation}
where
\begin{equation}
\begin{split}
\alpha_n(t;\varphi,\psi) = {}& \left|\sin\frac{\lambda t}{2}\right|
\frac{t}{n}\left|\langle \psi|A|\varphi\rangle\right| \notag \\
& - \frac{t^2}{4n^2}\min\Big\{\inf_{g\in\mathbb{R}}
\|(A-g)^2\varphi\|,\ \notag \\
&\qquad\ \inf_{g\in\mathbb{R}} \|(A-g)\varphi\|\,\|(A-g)\psi\|\Big\}
\notag \\
& - \frac{t^3}{24n^2}\|[A,[A,B]]\|_{\mathrm{op}}\\
& - \frac{t^3}{12n^2}\|[B,[B,A]]\|_{\mathrm{op}}.
\label{eq:6}
\end{split}
\end{equation}
For fixed $t>0$, this bound and the corresponding upper bound~\cite{Hahn2025} imply $\xi_n(t;\varphi)=\Theta(1/n)$ as $n\to\infty$, provided that there exists an eigenstate $\psi$ with $h\neq\kappa$ and $|\sin(\lambda t/2)|\,|\langle\psi|A|\varphi\rangle|>0$. When this coefficient vanishes, faster convergence is possible. Further, Hahn et al.~\cite{Hahn2025} show that under the same assumptions as in the state-dependent case, the norm Trotter error $b_n(t)$ defined as the following operator norm
\begin{equation}
b_n(t) \equiv \sup _{\|\psi\|=1}  \| ((e^{-i \frac{t}{n} A} e^{-i \frac{t} {n} B)})^n - e^{-i t (A+B)})\, \psi \|
\end{equation}
for the Hamiltonian $A+B$ can be bounded from
below by
\begin{equation}
b_n(t) \geq \max\{0,\ \sup_{\varphi,\psi} \alpha_n(t;\varphi,\psi),\
\beta_n(t)\},
\label{eq:7}
\end{equation}
where the supremum is taken over all pairs of normalized eigenstates
$\varphi,\psi$ of $A+B$ and
\begin{align}
\beta_n(t) = {}& \max\Bigg\{
\frac{t^2}{2n}\left(1 - \frac{t}{2n}\|A\|_{\mathrm{op}}\right)
Z_1(t),\ \notag \\
&\qquad\ \frac{t^2}{2n}\left(Z_1(t) - \frac{t}{2n}\|A\|_{\mathrm{op}}
Z_2(t)\right)
\Bigg\} \notag \\
& - \frac{t^3}{24n^2}\|[A,[A,B]]\|_{\mathrm{op}}
- \frac{t^3}{12n^2}\|[B,[B,A]]\|_{\mathrm{op}},
\label{eq:8}
\end{align}
for all $0 \leq t \leq 2n/\|A\|_{\mathrm{op}}$, with $Z_1(t)$ and
$Z_2(t)$ given in Eqs.~(9) and (10) in~\cite{Hahn2025} . For fixed $t>0$ with $Z_1(t)>0$, this bound and the upper bound (4) in~\cite{Hahn2025} imply $b_n(t)=\Theta(1/n)$ as $n\to\infty$.

\subsection{Quantum Localization and Error Suppression} 

In search of tighter Trotter error bounds in specific instances, Heyl et al.~\cite{sciadv.aau8342} demonstrate that Trotter error, when measured on local observables, is much smaller than what the worst-case bounds on the global wavefunction predict, and that this robustness is tied to a localization phenomenon in the eigenvector basis of the associated Floquet Hamiltonian operator. Specifically, consider a Hamiltonian $H$ decomposed into $M$ implementable terms $H=\sum_{l=1}^{M}H_l$, and the Trotterized evolution operator after $n$ repetitions of period $t/n$: 
\begin{equation}
U^{(n)}(t)=\Big[U_1\Big(\tfrac{t}{n}\Big)U_2\Big(\tfrac{t}{n}\Big)\cdots U_M\Big(\tfrac{t}{n}\Big)\Big]^{n}.
\end{equation}
For bounded Hermitian terms $H_l$, the discrepancy between the exact evolution $U(t)=e^{-itH}$ and $U^{(n)}(t)$ satisfies the operator-norm bound~\cite{PhysRevX.11.011020} 
\begin{equation}
\|U(t)-U^{(n)}(t)\|_{\mathrm{op}}\leq\frac{t^2}{2n}\sum_{l>m=1}^{M}\|[H_l,H_m]\|_{\mathrm{op}}.
\end{equation}
This bound grows quadratically with total time $t$ and, for the Ising chain considered below with fixed couplings, at most linearly with system size $N$. The authors recast the Trotterized dynamics as a periodically driven Floquet system with period $\tau=t/n$, defining the Floquet Hamiltonian $H_{\rm F}$ as follows: 
\begin{equation}
e^{-iH_{\rm F}\tau}=U_1(\tau)U_2(\tau)\cdots U_M(\tau).
\end{equation}
For small $\tau$, Magnus expansion gives
\begin{equation}
H_{\rm F}=H+i\frac{\tau}{2}\sum_{l>m}[H_l,H_m]+\mathcal{O}(\tau^2).
\end{equation}
 The authors~\cite{sciadv.aau8342} illustrate their discussion using the quantum Ising chain
\begin{equation}
\begin{split}
H =H_Z+H_X,\\
H_Z=J\sum_{l=1}^{N-1}S_l^zS_{l+1}^z+h\sum_{l=1}^{N}S_l^z,\\
H_X =g\sum_{l=1}^{N}S_l^x.
\end{split}
\end{equation}
as their benchmark problem, with $h/J=g/J=1$ and the two-gate Trotter sequence
\begin{equation}
U^{(1)}=U_1U_2,\ U_1=e^{-i\tau H_Z},\  U_2=e^{-i\tau H_X}.
\end{equation}
Next, the authors~\cite{sciadv.aau8342} consider inverse participation ratio (IPR) of the initial state in the Floquet eigenvector basis $\{|\phi_v\rangle\}$:
\begin{equation}
\mathrm{IPR}=\sum_v p_v^2,\ p_v=|\langle\phi_v|\psi_0\rangle|^2, 
\end{equation}
with the initial state $|\psi_0\rangle$, and out-of-time-ordered correlator (OOC), quantifying scrambling:
\begin{equation}
\begin{split}
\mathcal{F}(t)=\langle V^\dagger(t)W^\dagger V(t)W\rangle,\\
V=W=N^{-1}\sum_l S_l^z .
\end{split}
\end{equation}
Heyl et al.~\cite{sciadv.aau8342} note that a sharp threshold in Trotter step size $\tau$ separates the localized regime with small $\tau$ and controllable error, from the many-body quantum chaotic regime with large $\tau$ where $\mathrm{IPR}$ and $\mathcal{F}$ approach their fully delocalized values. In the finite systems and localized regimes studied, the authors find local-observable errors that remain controlled at long times and show no notable system-size dependence. They caution that eventual heating may require the step size to decrease with simulation time in the thermodynamic limit. Because local-observable errors in this regime remain controlled even for comparatively large Trotter steps, fewer gates are needed to reach the target simulation time, reducing exposure to extrinsic hardware errors in near-term digital quantum simulators.

\subsection{Low-Energy Subspace Simulations}

Similar to the Heyl et al.~\cite{sciadv.aau8342} study, focused on finding tighter Trotter error bounds in a specific computational regime, Şahinoğlu and Somma~\cite{sahinoglu2021} have analyzed Trotter errors when approximating evolution of spin systems with the initial state supported on the low energy subspace of the underlying Hamiltonian, demonstrating that in this instance Trotterization error depends on the effective low-energy norm of the Hamiltonian. The study is motivated by the fact that known numerical error bounds for product formulas depend on the norm of the Hamiltonian terms, and can be large even when the system stays in the low-energy states. Sahinoğlu and Somma~\cite{sahinoglu2021} derive Trotter error bounds that depend on the `effective low-energy norm' of the Hamiltonian, with tighter complexity estimates for long evolution times and high precision in the low energy subspace. Specifically, they consider the following $p$th-order product formula defined for the Hamiltonian $H = \sum_l H_l$ as
\begin{equation}
W_p(s) = e^{-i s_q H_{l_q}} \cdots e^{-i s_2 H_{l_2}} e^{-i s_1 H_{l_1}}
\end{equation}
and known general-case error bound 
\begin{equation}
\|U(s) - W_p(s)\| = \mathcal{O}((h|s|)^{p+1}), 
\end{equation}
where $h = \max_l \|H_l\|$. Let $\Delta$ be the energy associated with the initial state
$|\psi\rangle$, such that $\Pi_{\le \Delta} |\psi\rangle = |\psi\rangle$, where $\Pi_{\le \Delta}$ is a 
projector onto the subspace spanned by eigen-states of $H$ of energies (eigenvalues) of at most $\Lambda \leq 0$, 
and $H_l \ge 0$. Şahinoğlu and Somma~\cite{sahinoglu2021} formulate their main result as follows: 
\begin{equation}
\|(U(s) - W_p(s))\Pi_{\le \Delta}\| = \mathcal{O}((\Delta' s)^{p+1}),
\end{equation}
where effective norm is
\begin{equation}
\Delta' = \Delta + \beta_1 J \log(\beta_2/(J|s|)) + \beta_3 J^2 N |s|,
\end{equation}
with $J$ being interaction strength, $N$ system size, and positive constants $\beta_i$. This leads to the Trotter number
\begin{equation}
r' = \mathcal{O}\left( \frac{(|t|(\Delta + J))^{1 + \frac{1}{p}}}{\varepsilon^{\frac{1}{p}}} + \frac{(|t| J \sqrt{N})^{1 + \frac{1}{2p+1}}}{\varepsilon^{\frac{1}{2p+1}}} \right).
\end{equation}
 The authors compare this result against the best known general-case complexity estimate 
 \begin{equation}
 \mathcal{O}(\tau^{1+1/p} N^{1/p}/\varepsilon^{1/p}), 
 \end{equation}
 where $\tau = |t|J$. For $p=1$, at fixed $\Delta,J,t,\varepsilon$ in the constant-locality, constant-degree setting of Ref.~\cite{sahinoglu2021}, this gives an improvement of order $N^{1/3}$ in the Trotter-number bound.

 Motivated by an observation that for higher order product formulas Trotter error reduction is not evident when the initial state is restricted to a low-energy subspace, Mizuta and Kuwahara~\cite{Mizuta2025} provide further Trotter error analysis, focusing on higher order Trotterised Hamiltonian simulations. Specifically, they consider a $k$-local Hamiltonian $H$, characterized by maximal energy scale per site, or `extensiveness' $g$, defined as 
\begin{equation}
\sum_{X \subset \Lambda; X \ni i} \| h_{x}\| \leq g,\ \forall i \in \Lambda. 
\end{equation}
 The Hamiltonian is composed of positive-semidefinite terms, with the initial state $|\psi\rangle$ closed in the low-energy subspace, s.t. $\Pi_{\le\Delta}|\psi\rangle=|\psi\rangle$, where $\Pi_{\le\Delta}$ is a projector, and the  $p$-th order Trotter error is bounded by
 \begin{equation}
    \begin{split}
        \varepsilon_{p,\Delta}(t)\equiv\|(e^{-iHt}-T_p(t))\Pi_{\le\Delta}\|\in \\
\begin{cases}
\mathcal{O}\big((gt)^p\Delta' t+\epsilon\big), \ \Gamma\in\mathcal{O}(1)\\[4pt]
\mathcal{O}\big(\{gt\log(N/\epsilon)\}^p\Delta' t+\epsilon\big), \ \text{otherwise,}
\end{cases}
    \end{split}
\end{equation}
with $\Delta'=\Delta+\mathcal{O}(g\log(N/\epsilon))$. This is linear in the initial-state energy $\Delta$ and logarithmic in the system size $N$, improving exponentially in $N$ on previous bounds. The corresponding Trotter number needed to simulate time $t$ with the error $\varepsilon$ is
\begin{equation}
    \begin{split}
r\in
\begin{cases}
\mathcal{O}\!\left(gt\left(\dfrac{\Delta t+gt\log(N/\varepsilon)}{\varepsilon}\right)^{1/p}\right),\ \Gamma\in\mathcal{O}(1)\\[10pt]
\mathcal{O}\!\left(gt\log(N/\epsilon)\left(\dfrac{\Delta t+gt\log(N/\varepsilon)}{\varepsilon}\right)^{1/p}\right),\\ \text{otherwise,}
\end{cases}
\end{split}
\end{equation}
compared to $r\in\mathcal{O}(gt(Ngt/\varepsilon)^{1/p})$ for an arbitrary initial state. Next, the authors show that the error bound and the Trotter number are optimal in $\Delta$, as setting $\Delta=Ng$ recovers known optimal scaling for an arbitrary initial state. For $\Gamma=\mathcal{O}(1)$, where $\Gamma$ is the number of groups of mutually commuting local terms, the authors conclude that Trotterization is substantially more efficient for low-energy initial states across the entire regime $\Delta\in o(Ng)$. They note that $\Delta$ is defined relative to the shifted, positive-semidefinite terms $h_X\to h_X+\|h_X\|$ rather than the ground-state energy, and suggest extensions to qDRIFT, multi-product formulas, and time-dependent Trotterization as future research directions.

\section{Optimization Techniques}

\subsection{Term Ordering and Compilation}
\label{sec:term_ordering}

Motivated by the fact that Trotter error of a Hamiltonian evolution operator depends on the order in which its individual unitary terms are applied, Tranter et al.~\cite{Tranter2019} have investigated different Trotter term ordering strategies. They note that only non-commuting Hamiltonian evolution terms need to be Trotterized, while the terms that commute with all the terms in a Hamiltonian form a totally commuting set, and can be combined and simulated as an independent operator. They demonstrate this fact using an incompatibility graph coloring of a qubit Hamiltonian for a hydrogen molecule in the STO-3G basis. Further, the authors conduct a comprehensive empirical study of different term ordering strategies for more general Hamiltonians used in realistic simulations of systems involving large numbers of spin-orbitals.  

In more recent work, Tate et al. \cite{tate2026commutation} ran a comparative study of various Trotterization ordering strategies on 1D and 2D Heisenberg-style Hamiltonians. They introduced a new method, \emph{group-evolve}, where they (exactly) evolve entire color groups at a time within Trotterization. In terms of coloring the compatibility graph, they consider a variety of methods: a greedy approach, an exact approach using Gurobi (an integer programming solver), a ``localized" approach that groups Pauli terms that act on the same set of qubits, and lastly, a 3-color approach that groups Pauli strings by their Pauli type ($X$, $Y$, or $Z$); the last two coloring approaches require that the Hamiltonian satisfy certain properties. The empirical results by Tate et. al \cite{tate2026commutation} suggest that the best ordering strategy is highly dependent on the Trotter parameters and the Hamiltonian structure, but their group-evolve method did consistently well overall. Moreover, nearly all of the ordering strategies considered consistently outperformed what could typically be achieved by choosing a random ordering.

\subsection{Randomized Product Formulas}

Childs et al.~\cite{childs2019randomization} showed that simply permuting the summands of a product formula in random order yields provably stronger worst-case error bounds than the conventional deterministic-order analysis, at no additional gate cost. The qDRIFT approach of Campbell~\cite{PhysRevLett.123.070503} abandons a fixed term ordering entirely and instead samples the terms at random. For a Hamiltonian
\[
H = \sum_{\ell=1}^L h_\ell H_\ell
\]
with $\norm{H_\ell} = 1$ and $h_\ell > 0$, define $\lambda = \sum_{\ell} h_\ell$. The qDRIFT algorithm performs $N$ random steps, each sampling index $\ell$ with probability
\[
p_\ell = h_\ell/\lambda
\]
and applying $e^{-i \lambda t H_\ell/N}$. The rotation angle $\lambda t/N$ is the same for every sampled term; the weight $h_\ell$ enters only through the sampling probability, and the sampled circuits stochastically drift towards the target evolution.
\begin{algorithm}[h]
\caption{qDRIFT Randomized Evolution}
\label{alg:qdrift}
\small
\begin{algorithmic}[1]
\State \textbf{Input:} $H = \sum_{\ell=1}^L h_\ell H_\ell$, time $t$, precision $\epsilon$
\State $\lambda \gets \sum_{\ell=1}^L h_\ell$
\State $N \gets \lceil 2 \lambda^2 t^2/\epsilon \rceil$
\For{$n = 1$ to $N$}
    \State Sample $\ell$ with probability $p_\ell = h_\ell/\lambda$
    \State Apply $e^{-i \lambda t H_\ell/N}$
\EndFor
\end{algorithmic}
\end{algorithm}
Averaged over the samples, the random circuit defines a quantum channel whose diamond-norm distance from the ideal evolution satisfies $\epsilon \lesssim 2\lambda^2 t^2/N$, independent of $L$~\cite{PhysRevLett.123.070503}, although $L$-dependence can remain hidden in $\lambda \leq \Lambda L$, where $\Lambda = \max_\ell h_\ell$. For molecular Hamiltonians, which naturally satisfy $\lambda \ll \Lambda L$ and where ungrouped Pauli decomposition yields $L \sim N^4$ with $N$ orbitals, qDRIFT is therefore far cheaper than deterministic methods, whose costs scale polynomially with $L$. Campbell reports gate-count reductions between $306\times$ and $1591\times$ for small molecules at precision $10^{-3}$~\cite{PhysRevLett.123.070503}. However, since the qDRIFT gate count grows as $(\lambda t)^2/\epsilon$, higher-order deterministic formulas, whose costs approach linear scaling in $t$, eventually overtake qDRIFT for sufficiently long simulation times or sufficiently small approximation errors. The qDRIFT guarantee applies to the sample-averaged channel. Provided that a fresh random circuit is generated for each repetition of the experiment (repeated runs are needed anyway to estimate observables), realizing the averaged channel adds no sampling overhead beyond the usual shot-noise scaling~\cite{PhysRevLett.123.070503}. An individual fixed realization of the random circuit, in contrast, typically deviates from the ideal unitary by an amount closer to $\sqrt{\epsilon}$; rigorous concentration bounds for single realizations, including the additional system-size factors that arise for worst-case guarantees, are given by Chen et al.~\cite{chen2021concentration}.

qSWIFT extends randomization of the qDRIFT scheme to an arbitrary order~\cite{nakaji2024qswift}. The $K$-th order qSWIFT channel augments the qDRIFT channel with correction terms that cancel the low-order errors, which reduces the systematic error to
\[
\epsilon \in O\left(\left((\lambda t)^2/N\right)^K\right).
\]
The gate count is still independent of $L$, and precision $\epsilon$ now requires only $N \in O((\lambda t)^2/\epsilon^{1/K})$ gates. Since the correction terms do not constitute a physical (completely positive, trace-preserving) channel, they cannot be sampled directly. Instead, qSWIFT estimates expectation values of observables with \emph{swift circuits}, which interleave the sampled rotations with controlled-$H_\ell$ operations on a single ancilla qubit, and combines the measurement outcomes in classical post-processing. The improved scaling thus comes at the price of an ancilla qubit, controlled operations, and extra circuit repetitions to estimate the correction terms, so qSWIFT pays off mainly when high precision is required. To reach a systematic propagation error of $10^{-6}$, third-order qSWIFT needs roughly a thousand times fewer gates than qDRIFT~\cite{nakaji2024qswift}.

\subsection{Partially Randomized Product Formulas}

Deterministic product formulas converge rapidly with the step size but their cost grows with the number of terms $L$; qDRIFT-type methods are insensitive to $L$ but converge only at first order in $1/N$. Partially randomized, or \emph{composite}, schemes combine the two by splitting the Hamiltonian
\begin{equation}
H = H_{\text{det}} + H_{\text{rand}},
\end{equation}
where $H_{\text{det}}$ collects a small number $L_{\text{det}}$ of large-weight terms that are simulated with a deterministic product formula, while $H_{\text{rand}}$ collects the remaining long tail of small-weight terms, which is simulated by random sampling. Various realizations of this idea have been proposed~\cite{kivlichan2019phase, ouyang2020compilation, jin2021partially, rajput2022hybridized, kiss2023importance}; Hagan and Wiebe~\cite{hagan2023composite} developed a general framework for such composite channels; they prove diamond-norm error bounds and characterize the partitions for which the composite gate cost falls below that of either constituent method. Partial randomization is most effective when
\begin{equation}
\lambda_{\text{rand}} = \sum_{j \in \text{rand}} |h_j| \ll \lambda
\quad \text{and} \quad
L_{\text{det}} \ll L,
\end{equation}
that is, when a few strong terms carry most of the Hamiltonian weight, as is typical for electronic structure Hamiltonians in quantum chemistry.

Günther et al.~\cite{gunther2025randomized} recently refined the composite approach for ground-state energy estimation. In their scheme, a deterministic product formula of order $p$ is applied to the decomposition of $H$ into the individual terms of $H_{\text{det}}$ and the single term $H_{\text{rand}}$; each factor $e^{-i \delta H_{\text{rand}}}$ is then implemented with a randomized, unbiased linear-combination-of-unitaries expansion~\cite{wan2022randomized} built from sampled Pauli rotations and Pauli operators, where the rotation angle depends on the sampled Taylor order. In single-ancilla quantum phase estimation, the deterministic part biases the estimated energy and the randomized part damps the amplitude of the measured signal; the step size and the number of random rotations control both effects. Günther et al.~\cite{gunther2025randomized} show that estimating the ground-state energy to precision $\epsilon$ requires
\begin{equation}
O\left(L_{\text{det}}\, C^{1/p}\, \epsilon^{-1-1/p}\right)
\end{equation}
exponentials of the deterministic terms, together with
\begin{equation}
O\left(\lambda_{\text{rand}}^2\, \epsilon^{-2}\right)
\end{equation}
randomly sampled Pauli rotations, where $C$ is the Trotter constant governing the ground-state-energy bias of the deterministic formula for the chosen partition. Only the residual weight $\lambda_{\text{rand}}$, not the full weight $\lambda$, enters the randomized cost, so partial randomization improves on fully randomized phase estimation, though in practice by modest constant factors. Together with an error analysis that targets the energy bias rather than the full evolution error, this brings the resource estimates for quantum chemistry benchmarks down by orders of magnitude relative to earlier product-formula-based estimates; for hydrogen chains, the numerics suggest system-size scaling competitive with qubitization~\cite{gunther2025randomized}. A practical subtlety is that the Trotter constant $C$ itself depends on the partition, so that the optimal assignment of terms to $H_{\text{det}}$ and $H_{\text{rand}}$ becomes a nontrivial optimization problem.

\subsection{Multi-Product Formulas}

Multi-product formulas (MPF) approximate Hamiltonian evolution as linear combinations of Trotter–Suzuki, often higher order, product formulas
\begin{equation}
\tilde{U}(t) = \sum_{j=1}^M c_j S_p(t/r_j)^{r_j}. 
\end{equation}
The coefficients $\{c_j\}$ are chosen to minimize the approximation error~\cite{Faehrmann_2022}. MPF implementation for Hamiltonian simulation was introduced by Childs and Wiebe in~\cite{childs-wiebe-2012} and is based on the Linear Combination of Unitaries (LCU) combined with the amplitude amplification. This original implementation uses fixed predetermined coefficients, and exhibits improved Trotter error compared the non-MPF algorithms. Motivated by the need to reduce circuit depth in the MPF algorithms, Faehrmann et al.~\cite{Faehrmann_2022} have recently introduced a \emph{randomized sampling} framework for implementing multi-product formulas together with two new families of multi-product formulas, achieving a notably improved scaling. 

Zhuk, Robertson, and Bravyi~\cite{zhuk-24} have made further contributions towards designing practical multi-product formulas for the near-term quantum architectures. Their first contribution extends the theory of Trotter error with commutator scaling, due to Childs et al.\cite{childs-wiebe-2012}, from ordinary product formulas to MPFs. Their main result is the proof that MPFs achieve a nearly quadratic reduction in Trotter error compared to the base product formula without increasing circuit depth or qubit connectivity, and with only a constant-factor increase in the number of circuit repetitions. In order to prove this result they use a new integral representation of MPFs based on the Euler--Maclaurin summation formula, which allows to express MPF error via nested commutators. Additionally, Zhuk, Robertson, and Bravyi~\cite{zhuk-24} have introduced dynamic multi-product formulas, where the coefficients are time-dependent and chosen to minimize Frobenius-norm of the approximation of the time-evolved state projection error. Numerical experiments with the Heisenberg spin-chain benchmark with the second- and fourth-order base product formulas show that MPFs substantially reduce Trotter-step requirements for a fixed error tolerance, and that the proposed minimax MPF outperforms both the best single Trotter formula and well-conditioned static MPF, once realistic shot noise is included,  keeping the condition number low. 

\subsection{Adaptive Trotterization}

Adaptive Trotterization~\cite{adaptive-trotter-2024} is another novel approach to finding an optimal balance between keeping Trotter error small, while optimizing the circuit depth. The authors had previously developed an adaptive step-size algorithm (ADA-Trotter) for time-independent Hamiltonians, which uses energy conservation as a diagnostic for time step $\delta t$. Extending this approach to explicitly time-dependent Hamiltonians $H(t)$ is nontrivial, since energy is no longer conserved and there is no fixed reference state against which to benchmark simulation accuracy. The authors address this problem defining a \emph{piecewise conserved} effective Hamiltonian, such that over a short time interval  exact evolution is equivalent to that generated by a single, locally static Hamiltonian $H_{[\infty]}$. The expectation and the higher moments of the Hamiltonian must match at the start and at the end of the interval if the step is simulated exactly. This effective Hamiltonian is approximated by a truncated series expansion, giving a computable quantity $H_{[k]}$ which expectation value and variance can be measured before and after each Trotter step. Tested on a driven  nonintegrable Ising spin chain, this globally constrained adaptive method suppresses late-time drift seen under the purely local constraint, keeps simulated magnetization accurate over much longer times, and outperforms a conventional fixed-step Trotter scheme at equal circuit depth, automatically reducing step size during strong driving and increasing it when driving is weak. The scheme reduces to the authors' earlier time-independent algorithm once the Hamiltonian becomes static.

\subsection{Product Formulas with Multiple Energy Scales}

Following research of Low and Wiebe~\cite{Low-Wiebe-2019}, who have introduced an efficient algorithm for time independent Hamiltonian simulation consisting of the time independent and time dependent Hamiltonian terms, Bosse et al.~\cite{Bosse2025} recently introduced a family of quantum algorithms - Trotter Heuristic Resource Improved Formulas for Time-dynamics (THRIFT) - for simulating time evolution of a quantum system with a 
Hamiltonian $H$ that naturally exhibits  multiple energy scales across its two additive terms - an exactly implementable 
term $H_0$ and a small perturbative term $\alpha H_1$: 
\begin{equation}
    H = H_0 + \alpha H_1, \qquad \alpha \ll 1,
\end{equation}
where $H_0$ and $H_1$ have comparable norms, and $e^{-itH_0}$ can be implemented exactly as a sequence of quantum gates 
for any $t$ with low resource requirements.

The basic THRIFT formula splits the perturbation into pieces and combines each piece with $H_0$. For example, if $H_1=A+B$, one THRIFT step approximates $e^{-itH}$ by
\[
U_{\mathrm{THRIFT}}(t)=
e^{-it(H_0+\alpha A)}\,e^{+itH_0}\,e^{-it(H_0+\alpha B)}.
\]
Each combined evolution therefore involves $H_0$ and only one piece of the perturbation, which can be simpler to implement than evolution under the full Hamiltonian. Realizing a resource advantage with basic THRIFT depends on implementing these combined evolutions inexpensively and with errors small enough to preserve the improved error scaling. For time-dependent Hamiltonians, the corresponding time-ordered evolutions are needed. Under these implementation assumptions, Bosse et al.~\cite{Bosse2025} show that THRIFT can improve the gate-complexity and circuit-depth scaling over a naive Trotter implementation.
By applying Trotterization to the perturbation in the interaction picture, the basic formula achieves
leading-order simulation error $O(\alpha^2 t^2)$ for \emph{sufficiently small $t$}, an improvement by a factor of $\alpha$ over the standard 
first-order Trotter error of $O(\alpha t^2)$. This approach results in an explicit product-formula based circuit
built with no ancilla qubits. 

Bosse et al.~\cite{Bosse2025} present the following versions of the THRIFT algorithm. The first, `THRIFT of order $k$' version, requires 
standard techniques for building higher-order product formulas from the first-order seed with error $O(\alpha^2 t^{k+1})$, 
versus $O(\alpha t^{k+1})$ for ordinary $k$th-order Trotter formulas. In contrast, `Magnus-THRIFT' and `Fer-THRIFT' variants are based on the Magnus and the Fer expansions of the time-ordered interaction-picture operators respectively.  These THRIFT versions achieve improved, $O(\alpha^{k+1} t^{k+1})$ scaling, though at a higher per-step implementation cost. The authors note that no product formula built purely from time-ordered evolutions under the Hamiltonian summands can beat the $\alpha^2$ scaling of the basic THRIFT. While asymptotic scaling of the algorithms in the THRIFT family is worse than that of such optimal methods as the Quantum Signal Processing, THRIFT implementations do not require either ancilla qubits or block encoding, and can be practical on near-term hardware when their constituent operations can be implemented efficiently. Additionally, THRIFT empirically outperforms standard product formulas in several models with separated energy scales, but this is not universal: implementation overhead can favor an optimized standard product formula, as in the Fermi--Hubbard benchmark. 

\section{Summary}

Modern product formula research is focused on designing algorithms with improved approximation accuracy that retain minimal ancilla overheads and optimal quantum circuit configurations, suitable for the near term gate-based hardware. Trotter error analysis across many distinct studies consistently indicates that rigorous worst-case error bounds are often pessimistic. 
The choice of a specific product formula approach often depends on the choice of Trotterization parameters, while in the application specific instances it often depends on the structure of the Hamiltonian and on the choice of the initial state. Trotter error can be minimized through the analysis of nested commutators of the Hamiltonian terms. Structure- and state-specific results provide further algorithmic improvements. In the localized regimes studied, errors on local observables can remain small for comparatively large steps, while low-energy subspace bounds can improve on worst-case estimates when their assumptions hold. At the same time, several complementary numerical optimization strategies further improve product formula performance. The order in which unitary terms are applied affects Trotter error, and structured orderings, including exact evolution of the entire color groups of commuting terms may outperform random orderings. In the randomized methods, such as the qDRIFT, and its higher-order extension, qSWIFT, gate counts are independent of the number of the Trotter terms. qSWIFT's higher-order accuracy costs an ancilla qubit, controlled operations and extra circuit repetitions. Partially randomized schemes treat the dominant terms deterministically and sample the long tail of small terms. Multi-product formulas implemented through classical combinations of observable estimates can reduce Trotter error nearly quadratically without increasing circuit depth, at the expense of a constant-factor increase in circuit repetitions. Adaptive step-size selection tunes the Trotter step for time-dependent Hamiltonians, and the THRIFT family exploits well-separated energy scales to improve on the standard first-order error without ancilla qubits, subject to its implementation requirements. 

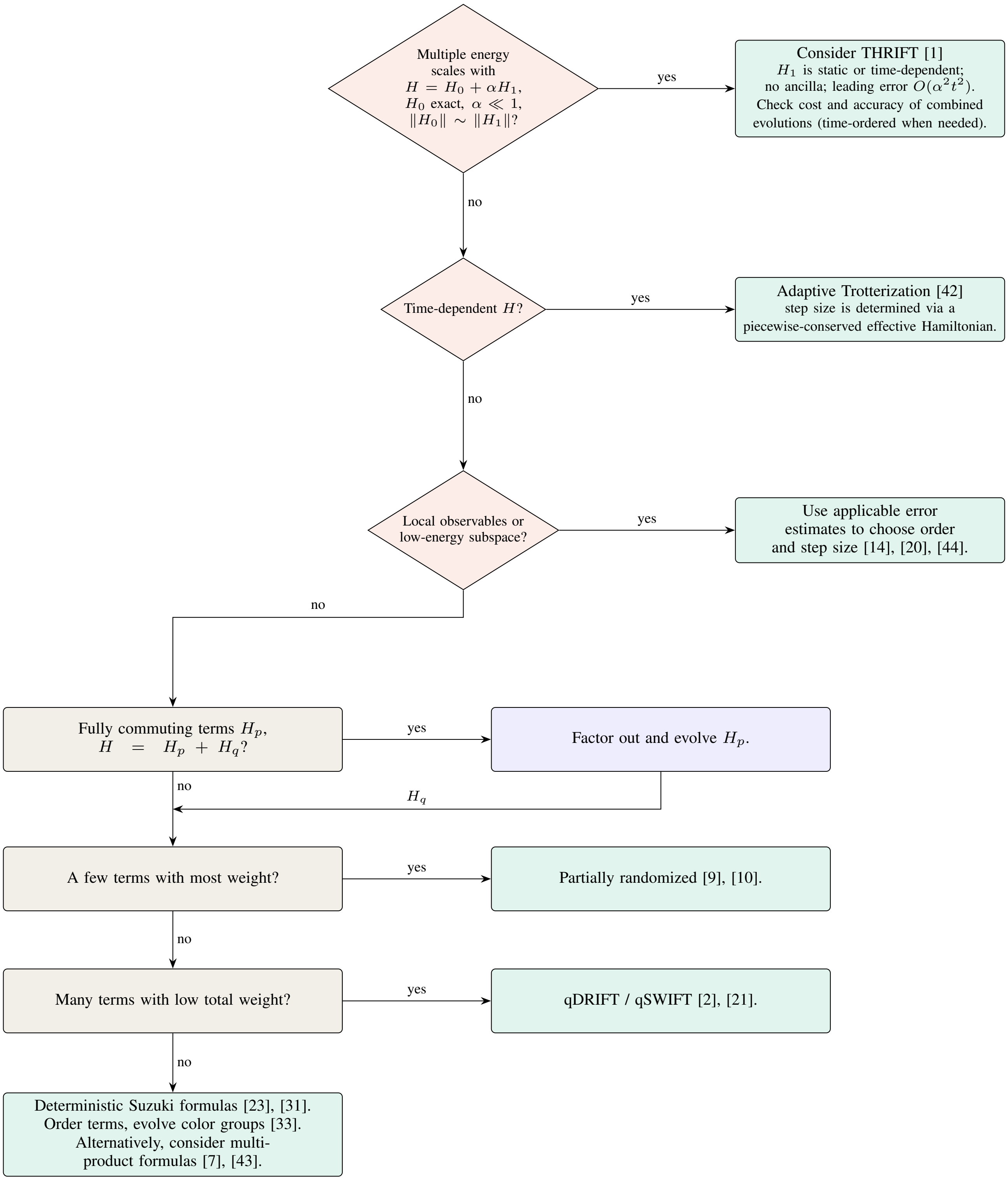
\begin{figure*}[t]
\centering
\resizebox{\textwidth}{!}{%
\begin{tikzpicture}[node distance=0pt]
 
  \node[decision] (d0) at (5, 11.2) {Multiple energy scales with\\
       {\scriptsize $H=H_0+\alpha H_1$,\\$H_0$ exact,\, $\alpha\ll1$, $\|H_0\| \sim \|H_1\|$?}};
  \node[tdleaf]   (l0) at (12, 11.2) {Consider THRIFT~\cite{Bosse2025}\\
       {\scriptsize $H_1$ is static or time-dependent;\\
        no ancilla; leading error $O(\alpha^2t^2)$.\\ Check cost and accuracy of combined evolutions (time-ordered when needed).}};
 
  \node[decision] (d1) at (5,7.4) {Time-dependent $H$?};
  \node[tdleaf]   (l1) at (12,7.4) {Adaptive Trotterization~\cite{adaptive-trotter-2024}\\
       {\scriptsize step size is determined via a\\
        piecewise-conserved effective Hamiltonian.}};
 
  \node[decision] (d2) at (5,3.6) {Local observables or low-energy subspace?};
  \node[tdleaf]   (l2) at (12,3.6) {Use applicable error estimates to choose order and step size~\cite{sciadv.aau8342, sahinoglu2021, Mizuta2025}.};

  \draw[flow] (d0.east) -- node[lab, above] {yes} (l0.west);
  \draw[flow] (d0.south) -- node[lab, right, pos=0.35] {no} (d1.north);
 
  \draw[flow] (d1.east) -- node[lab, above] {yes} (l1.west);
  \draw[flow] (d1.south) -- node[lab, right, pos=0.35] {no} (d2.north);
  \draw[flow] (d2.east) -- node[lab, above] {yes} (l2.west);

  \node[q]   (q0) at (0,0) {Fully commuting terms $H_{p}$,\\ $H=H_p + H_q$?};
  \node[pre] (p)  at (8.4,0) {Factor out and evolve $H_p$. };

  \node[q]   (q4) at (0,-2.4)  {A few terms with most weight?};  
  \node[leaf](l4) at (8.4,-2.4){Partially randomized~\cite{hagan2023composite, gunther2025randomized}.\\
       %{\scriptsize \emph{deterministic large terms, sampled tail}}
       };

  \node[q]   (q5) at (0,-4.5)   {Many terms with low total weight?};
  \node[leaf](l5) at (8.4,-4.5) {qDRIFT / qSWIFT~\cite{PhysRevLett.123.070503, nakaji2024qswift}.\\
       %{\scriptsize \emph{qSWIFT: high precision, one ancilla}}
       };
  
  \node[leaf](f)  at (0,-6.8) {Deterministic Suzuki formulas~\cite{Suzuki1991, Ostmeyer_2023}.\\
       {Order terms, evolve color groups~\cite{tate2026commutation}.\\} 
       Alternatively, consider multi-product formulas~\cite{Faehrmann_2022, zhuk-24}. 
       %{\scriptsize \emph{(lower error, increased circuit repetitions)}}
  };
 
  \coordinate (mj) at ($(q0.south)!0.5!(q4.north)$);
  \draw[flow] (q0.east) -- node[lab, above] {yes} (p.west);
  \draw[flow] (q0.south) -- node[lab, right, pos=0.2] {no} (q4.north);
  \draw[flow] (p.south) |- node[lab, above, pos=0.75] {$H_q$} (mj);
 
  \foreach \a/\b in {q4/l4, q5/l5}
    \draw[flow] (\a.east) -- node[lab, above] {yes} (\b.west);
  \foreach \a/\b in {q4/q5, q5/f}
    \draw[flow] (\a.south) -- node[lab, right] {no} (\b.north);
 
  \draw[flow] (d2.south) -- (5,2.1) -- (0,2.1) -- (q0.north);
  \node[lab] at (2.5,2.3) {no};
 
\end{tikzpicture}}
\caption{This flow chart offers a possible product formula selection strategy based on the methods surveyed in the paper.}

\label{fig:tree}
\end{figure*}

Fig.~\ref{fig:tree} is an attempt to summarize our study as a flow chart with suggestions for selecting the most 
appropriate Trotterization scheme for a given quantum simulation problem. We first look at the application specific scenarios. Specifically, we check if a Hamiltonian exhibits multiple energy scales with a perturbative structure and efficiently implementable combined evolutions of sufficient accuracy to consider THRIFT~\cite{Bosse2025}, for either static or time-dependent Hamiltonians. Alternatively, we check if the Hamiltonian is time-dependent, in which case adaptive  Trotterization~\cite{adaptive-trotter-2024} is suggested. Since energy is no longer conserved in the time-dependent instance, adaptive Trotter computes piecewise-conserved  effective Hamiltonian to set the step size~\cite{adaptive-trotter-2024}. Next, we check whether applicable results for local observables or a low-energy subspace permit improved error estimates, and use those estimates to choose the Trotter order and step size~\cite{sciadv.aau8342, sahinoglu2021, Mizuta2025}. If the application specific information about the Hamiltonian is not available,  we move on to the numerical optimization strategies, suggesting to first factor out and evolve any exactly implementable fully commuting terms $H_p$. The terms that commute with all other terms form a totally commuting set that can be combined and simulated as an independent operator, so only the non-commuting remainder $H_q$ needs to be Trotterized. Next, in the more general Hamiltonian instances, randomized~\cite{PhysRevLett.123.070503, nakaji2024qswift} and partially randomized~\cite{hagan2023composite, gunther2025randomized} product formulas, classical formulas~\cite{Suzuki1991, Ostmeyer_2023}, formulas with term reordering~\cite{tate2026commutation}, and the multi-product formulas should be considered~\cite{Faehrmann_2022, zhuk-24}. 

 \section{Acknowledgments}
This work had been supported by the New Mexico Consortium Quantum Frontier Project, and it had been assigned Los Alamos National Laboratory unrestricted release number LA-UR-26-24076. Claude Sonnet 4.6 and Sonnet 5 Medium had been used during the preparation of the paper to help us summarize various cited results, to typeset the formulas, and to help us design and plot the flow chart. We would like to thank Yigit Subasi and Brendan Krueger for their valuable suggestions and feedback throughout the preparation of the manuscript.
 
\bibliographystyle{plain}
\bibliography{trotter_refs}

@article{Spiteri-Wei-23,
title = {Fractional-step {R}unge–{K}utta methods: representation and linear stability analysis},
journal = {Journal of {C}omputational {P}hysics},
volume = {476},
pages = {111900},
year = {2023},
issn = {0021-9991},
doi = {https://doi.org/10.1016/j.jcp.2022.111900},
url = {https://www.sciencedirect.com/science/article/pii/S0021999122009639},
author = {Raymond J. Spiteri and Siqi Wei}
}

@misc{Low-Wiebe-2019,
      title={Hamiltonian Simulation in the Interaction Picture}, 
      author={Guang Hao Low and Nathan Wiebe},
      year={2019},
      eprint={1805.00675},
      archivePrefix={arXiv},
      primaryClass={quant-ph},
      url={https://arxiv.org/abs/1805.00675}
}

@article{childs-wiebe-2012, 
   author={Andrew M. Childs and Nathan Wiebe},
   title={Hamiltonian Simulation Using Linear Combinations of Unitary Operations},
   volume={12},
   ISSN={1533-7146},
   url={http://dx.doi.org/10.26421/QIC12.11-12},
   DOI={10.26421/qic12.11-12},
   journal={Quantum Information and Computation},
   publisher={Rinton Press},
   year={2012},
   month=Nov
   }

@article{Faehrmann_2022,
   title={Randomizing multi-product formulas for {H}amiltonian simulation},
   volume={6},
   ISSN={2521-327X},
   url={http://dx.doi.org/10.22331/q-2022-09-19-806},
   DOI={10.22331/q-2022-09-19-806},
   journal={Quantum},
   publisher={Verein zur Forderung des Open Access Publizierens in den Quantenwissenschaften},
   author={Faehrmann, Paul K. and Steudtner, Mark and Kueng, Richard and Kieferova, Maria and Eisert, Jens},
   year={2022},
   pages={806}
   }

@misc{tate2026commutation,
      title={An Analysis of Commutation-Based {T}rotter Ordering Strategies on {H}eisenberg-Style {H}amiltonians}, 
      author={Reuben Tate and Shamminuj Aktar and Stephan Eidenbenz},
      year={2026},
      eprint={2604.23138},
      archivePrefix={arXiv},
      primaryClass={quant-ph},
      url={https://arxiv.org/abs/2604.23138}, 
}

@book{yanenko-2012,
   title={The Method of Fractional Steps}, 
   subtitle={The Solution of Problems of Mathematical Physics in Several Variables}, 
   publisher={Springer Berlin, Heidelberg}, 
   author={N.N. Yanenko}, 
   year={2012},
   note={Copyright Information: Springer-Verlag, Berlin Heidelberg 1971},
   url={https://link.springer.com/book/10.1007/978-3-642-65108-3}
}

@article{Lie-1888,
   title={Theorie der {T}ransformationsgruppen},
   publisher={Leipzig: {BG} {T}eubner},
   author={S. Lie},
   year={1888-1893},
url={https://archive.org/details/theotransformation01liesrich},
note={Contributor: {U}niversity of {C}alifornia {L}ibraries}
}

@article{Peaceman-Rachford-55,
author = {Peaceman, D. W. and Rachford, Jr., H. H.},
title = {The Numerical Solution of Parabolic and Elliptic Differential Equations},
journal = {Journal of the {S}ociety for {I}ndustrial and {A}pplied {M}athematics},
volume = {3},
number = {1},
pages = {28-41},
year = {1955},
doi = {10.1137/0103003},
URL = {https://doi.org/10.1137/0103003},
eprint = {https://doi.org/10.1137/0103003}
}

@article{Trotter-58,
author = {Trotter, H.F.}, 
title = {Approximation of {S}emi-{G}roups of {O}perators}, 
journal = {{P}acific {J}ournal of {M}athematics},
number = 8,
pages = {887-919},
URL = {http://dx.doi.org/10.2140/pjm.1958.8.887},
year = 1958
}

@article{Trotter-1959,
   title={{O}n the {P}roduct of {S}emi-{G}roups of {O}perators},
   volume={10},
   url={http://dx.doi.org/10.1090/S0002-9939-1959-0108732-6},
   pages={545-551},
   journal={Proceedings of the {A}merican {M}athematical {S}ociety},
   publisher={American {M}athematical {S}ociety},
   author={H. F. Trotter},
   year={1959}
}

@article{godunov-59,
author = {S. K. Godunov},
title = {A difference method for numerical calculation of discontinuous solutions of the equations of hydrodynamics},
journal = {{M}atematicheskii {S}bornik},
volume = 47,
number = 89,
pages = "271--306",
year = 1959,
note = {In {R}ussian}
}

@article{yanenko-marchuk-66,
author = {N.N. Yanenko and G.I. Marchuk},
title = {Application of the Method of Splitting (Fractional Steps) to the Solution of Problems of mathematical Physics}, 
journal = {Certain Questions of Computational and Applied Mathematics},
pages = {5-22}, 
year = 1966,
place={Novosibirsk},
note = {In {R}ussian}
}

@article{strang-68,
author = {Strang, Gilbert},
title = {On the construction and comparison of difference schemes},
journal = {{SIAM} Journal on {N}umerical {A}nalysis},
number = {5}, 
pages = {506–517},
year = 1968,
url={https://doi.org/10.1137/0705041}
}

@article{suzuki-75,
Author= {Suzuki, Masuo},
Title={Generalized {T}rotter's formula and systematic approximants of exponential operators and inner derivations with applications to many-body problems},
Journal={Communications in {M}athematical {P}hysics},
Year=1976,
Volume={51},
Number={2},
Pages= {183-190},
}

@article{suzuku-85,
    author = {Suzuki, Masuo},
    title = {Decomposition {F}ormulas of {E}xponential {O}perators and {L}ie {E}xponentials with {S}ome {A}pplications to {Q}uantum {M}echanics and {S}tatistical {P}hysics},
    journal = {{J}ournal of {M}athematical {P}hysics},
    volume = {26},
    number = {4},
    pages = {601-612},
    year = {1985},
    month = {04},
    issn = {0022-2488},
    doi = {10.1063/1.526596},
    url = {https://doi.org/10.1063/1.526596}
}

@article{suzuki-92,
title = {General theory of higher-order decomposition of exponential operators and symplectic integrators},
journal = {{P}hysics {L}etters {A}},
volume = {165},
number = {5},
pages = {387-395},
year = {1992},
issn = {0375-9601},
doi = {https://doi.org/10.1016/0375-9601(92)90335-J},
url = {https://www.sciencedirect.com/science/article/pii/037596019290335J},
author = {Masuo Suzuki}
}

@book{yanenko-91,
   place={Moscow},
   title={Selected Works},
   publisher={Nauka},
   author={Yanenko, N.N},
   year=1991,
   note = {In {R}ussian},
}

@Inbook{Hatano2005,
author="Hatano, Naomichi and Suzuki, Masuo",
title="Finding Exponential Product Formulas of Higher Orders",
bookTitle="Quantum Annealing and Other Optimization Methods",
year="2005",
publisher="Springer {B}erlin {H}eidelberg",
address="Berlin, Heidelberg",
pages="37--68",
isbn="978-3-540-31515-5",
doi="10.1007/11526216_2",
url="https://doi.org/10.1007/11526216_2"
}

@article{Ostmeyer_2023,
doi = {10.1088/1751-8121/acde7a},
url = {https://doi.org/10.1088/1751-8121/acde7a},
year = {2023},
month = {Jun},
publisher = {IOP Publishing},
volume = {56},
number = {28},
pages = {285303},
author = {Ostmeyer, Johann},
title = {Optimised {T}rotter decompositions for classical and quantum computing},
journal = {Journal of {P}hysics {A}: {M}athematical and {T}heoretical}
}

@article{wei-guenter-25,
author = {Wei, Siqi and Guenter, Victoria and Spiteri Raymond J.},
title = {Improving the Stability and Efficiency of Operator-Splitting Methods},
journal = {{SIAM} {J}ournal on {S}cientific {C}omputing},
volume = {47},
number = {3},
pages = {A1888-A1906},
year = {2025},
doi = {10.1137/24M1683184},
URL = {https://doi.org/10.1137/24M1683184},
eprint = {https://doi.org/10.1137/24M1683184}
}

@article{Suzuki1991,
  author = {Suzuki, Masuo},
  title = {General theory of fractal path integrals with applications to many-body theories and statistical physics},
  journal = {Journal of Mathematical Physics},
  volume = {32},
  number = {2},
  pages = {400--407},
  year = {1991},
  doi = {10.1063/1.529425}
}

@article{Tranter2019,
  author = {Tranter, Andrew and Love, Peter J. and Mintert, Florian and Wiebe, Nathan and Coveney, Peter V.},
  title = {Ordering of {T}rotterization: Impact on Errors in Quantum Simulation of Electronic Structure},
  journal = {Entropy},
  volume = {21},
  number = {12},
  pages = {1218},
  year = {2019},
  doi = {10.3390/e21121218}
}

@article{adaptive-trotter-2024,
  author = {Zhao, Hongzheng and Bukov, Marin and Heyl, Markus and Moessner, Roderich},
  title = {Adaptive Trotterization for Time-Dependent {H}amiltonian Quantum Dynamics Using Piecewise Conservation Laws},
  journal = {Physical Review Letters},
  volume = {133},
  number = {1},
  pages = {010603},
  year = {2024},
  doi = {10.1103/PhysRevLett.133.010603}
}

@article{sciadv.aau8342,
  author = {Heyl, Markus and Hauke, Philipp and Zoller, Peter},
  title = {Quantum localization bounds {T}rotter errors in digital quantum simulation},
  journal = {Science Advances},
  volume = {5},
  number = {4},
  year = {2019},
  doi = {10.1126/sciadv.aau8342}
}

@article{PhysRevX.11.011020,
  author = {Childs, Andrew M. and Su, Yuan and Tran, Minh C. and Wiebe, Nathan and Zhu, Shuchen},
  title = {Theory of {T}rotter Error with Commutator Scaling},
  journal = {Physical Review X},
  volume = {11},
  number = {1},
  pages = {011020},
  year = {2021},
  doi = {10.1103/PhysRevX.11.011020}
}

@article{Bosse2025,
  author = {Bosse, Jan Lukas and Childs, Andrew M. and Derby, Charles and Gambetta, Filippo Maria and Montanaro, Ashley and Santos, Raul A.},
  title = {Efficient and practical Hamiltonian simulation from time-dependent product formulas},
  journal = {Nature Communications},
  volume = {16},
  number = {1},
  pages = {2673},
  year = {2025},
  doi = {10.1038/s41467-025-57580-5}
}

@article{Hahn2025,
  author = {Hahn, Alexander and Hartung, Paul and Burgarth, Daniel and Facchi, Paolo and Yuasa, Kazuya},
  title = {Lower bounds for the {T}rotter error},
  journal = {Physical Review A},
  volume = {111},
  number = {2},
  pages = {022417},
  year = {2025},
  doi = {10.1103/PhysRevA.111.022417}
}

@article{gunther2025randomized,
  author = {Günther, Jakob and Witteveen, Freek and Schmidhuber, Alexander and Miller, Marek and Christandl, Matthias and Harrow, Aram},
  title = {Phase estimation with partially randomized time evolution},
  journal = {PRX Quantum},
  volume = {7},
  pages = {020332},
  year = {2026},
  doi = {10.1103/ynxb-p2xq},
  eprint = {2503.05647},
  archivePrefix = {arXiv},
  primaryClass = {quant-ph}
}

@book{Nielsen-Chuang,
  author = {Nielsen, Michael A. and Chuang, Isaac L.},
  title = {Quantum Computation and Quantum Information},
  publisher = {Cambridge University Press},
  year = {2010}
}

@book{Hall-LieAlgebras,
  author = {Hall, Brian C.},
  title = {{L}ie Groups, {L}ie Algebras, and Representations: An Elementary Introduction},
  publisher = {Springer},
  year = {2015}
}

@article{sahinoglu2021,
  author = {Şahinoğlu, Burak and Somma, Rolando D.},
  title = {Hamiltonian simulation in the low-energy subspace},
  journal = {npj Quantum Information},
  volume = {7},
  number = {1},
  year = {2021},
  month = jul,
  doi = {10.1038/s41534-021-00451-w},
  publisher = {Springer Science and Business Media LLC}
}

@article{Mizuta2025,
  title = {Trotterization is Substantially Efficient for Low-Energy States},
  author = {Mizuta, Kaoru and Kuwahara, Tomotaka},
  journal = {Phys. Rev. Lett.},
  volume = {135},
  issue = {13},
  pages = {130602},
  numpages = {6},
  year = {2025},
  month = {Sep},
  publisher = {American Physical Society},
  doi = {10.1103/q87n-5xhz},
  url = {https://link.aps.org/doi/10.1103/q87n-5xhz}
}

@article{PhysRevLett.123.070503,
  author = {Campbell, Earl},
  title = {Random Compiler for Fast {H}amiltonian Simulation},
  journal = {Physical Review Letters},
  volume = {123},
  number = {7},
  pages = {070503},
  year = {2019},
  month = {Aug},
  publisher = {American Physical Society},
  doi = {10.1103/PhysRevLett.123.070503}
}

@article{zhuk-24,
  title = {Trotter error bounds and dynamic multi-product formulas for {H}amiltonian simulation},
  author = {Zhuk, Sergiy and Robertson, Niall F. and Bravyi, Sergey},
  journal = {PHYSICAL REVIEW RESEARCH},
  volume = {6},
  year = {2024},
  publisher = {American Physical Society},
  url = {https://journals.aps.org/prresearch/pdf/10.1103/PhysRevResearch.6.033309}
}

@article{childs2019randomization,
  author = {Childs, Andrew M. and Ostrander, Aaron and Su, Yuan},
  title = {Faster quantum simulation by randomization},
  journal = {Quantum},
  volume = {3},
  pages = {182},
  year = {2019},
  doi = {10.22331/q-2019-09-02-182}
}

@article{chen2021concentration,
  author = {Chen, Chi-Fang and Huang, Hsin-Yuan and Kueng, Richard and Tropp, Joel A.},
  title = {Concentration for random product formulas},
  journal = {PRX Quantum},
  volume = {2},
  number = {4},
  pages = {040305},
  year = {2021},
  doi = {10.1103/PRXQuantum.2.040305}
}

@article{nakaji2024qswift,
  author = {Nakaji, Kouhei and Bagherimehrab, Mohsen and Aspuru-Guzik, Alán},
  title = {{qSWIFT}: High-order randomized compiler for {H}amiltonian simulation},
  journal = {PRX Quantum},
  volume = {5},
  number = {2},
  pages = {020330},
  year = {2024},
  doi = {10.1103/PRXQuantum.5.020330}
}

@article{ouyang2020compilation,
  author = {Ouyang, Yingkai and White, David R. and Campbell, Earl T.},
  title = {Compilation by stochastic {H}amiltonian sparsification},
  journal = {Quantum},
  volume = {4},
  pages = {235},
  year = {2020},
  doi = {10.22331/q-2020-02-27-235}
}

@article{hagan2023composite,
  author = {Hagan, Matthew and Wiebe, Nathan},
  title = {Composite quantum simulations},
  journal = {Quantum},
  volume = {7},
  pages = {1181},
  year = {2023},
  doi = {10.22331/q-2023-11-14-1181}
}

@article{rajput2022hybridized,
  author = {Rajput, Abhishek and Roggero, Alessandro and Wiebe, Nathan},
  title = {Hybridized methods for quantum simulation in the interaction picture},
  journal = {Quantum},
  volume = {6},
  pages = {780},
  year = {2022},
  doi = {10.22331/q-2022-08-17-780}
}

@misc{jin2021partially,
  author = {Jin, Shi and Li, Xiantao},
  title = {A partially random {T}rotter algorithm for quantum {H}amiltonian simulations},
  year = {2021},
  note = {arXiv:2109.07987},
  eprint = {2109.07987},
  archivePrefix = {arXiv},
  primaryClass = {quant-ph}
}

@misc{kivlichan2019phase,
  author = {Kivlichan, Ian D. and Granade, Christopher E. and Wiebe, Nathan},
  title = {Phase estimation with randomized {H}amiltonians},
  year = {2019},
  note = {arXiv:1907.10070},
  eprint = {1907.10070},
  archivePrefix = {arXiv},
  primaryClass = {quant-ph}
}

@article{wan2022randomized,
  author = {Wan, Kianna and Berta, Mario and Campbell, Earl T.},
  title = {Randomized quantum algorithm for statistical phase estimation},
  journal = {Physical Review Letters},
  volume = {129},
  number = {3},
  pages = {030503},
  year = {2022},
  doi = {10.1103/PhysRevLett.129.030503}
}

@article{kiss2023importance,
  author = {Kiss, Oriel and Grossi, Michele and Roggero, Alessandro},
  title = {Importance sampling for stochastic quantum simulations},
  journal = {Quantum},
  volume = {7},
  pages = {977},
  year = {2023},
  doi = {10.22331/q-2023-04-13-977}
}
\end{document}